\documentclass[12pt,letterpaper]{article}
\pdfoutput=1

\usepackage{graphicx,array}
\usepackage[dvipsnames]{xcolor}
\definecolor{darkblue}{rgb}{0,0.1,0.5}
\definecolor{darkgreen}{rgb}{0,0.5,0.2}
\definecolor{darkred}{RGB}{153,26,0}
\usepackage{ulem}
\definecolor{seablue}{rgb}{0,0.2,0.6}
\usepackage{latexsym}
\usepackage{amssymb, amsmath}
\usepackage{bm}        
\usepackage[numbers,sort&compress]{natbib}
\usepackage{bm,relsize}
\usepackage{slashed}
\usepackage{tikzfeynman}
\definecolor{viola}{RGB}{134,41,198}
\usepackage{mathrsfs}
\usepackage{hyperref} 
\hypersetup{
    colorlinks=true,       
    linkcolor=darkblue,          
    citecolor=BrickRed,        
    filecolor=magenta,      
    urlcolor=MidnightBlue           
}
\usepackage[all]{hypcap} 
\usepackage{subcaption}

\usepackage[font=small,labelfont=bf,labelsep=period]{caption}

\newcommand{\Mpl}{M_{\rm Pl}}
\newcommand{\op}{\mathcal{O}}

\newcommand{\be}{\begin{equation}}
\newcommand{\ee}{\end{equation}}

 \date{\today}

\begin{document}

\begin{flushright}

\end{flushright}
\vspace{.6cm}
\begin{center}
{\LARGE \bf General Freeze-in and Freeze-out
}\\
\bigskip\vspace{1cm}
{
\large Michele Redi, Andrea Tesi}
\\[7mm]
 {\it \small
INFN Sezione di Firenze, Via G. Sansone 1, I-50019 Sesto Fiorentino, Italy\\
Department of Physics and Astronomy, University of Florence, Italy
 }

\vspace{2cm}

\centerline{\bf Abstract} 
\begin{quote}
We use the framework of relativistic and non-relativistic conformal field theories (CFT) to derive general results relevant
for the production of weakly coupled and strongly coupled dark sectors  through thermal interactions. Our result reproduce trivially known formulas for $2\to n $ processes and extend to general $m\to n$ processes as well as interacting dark sectors. As concrete examples we consider freeze-in of a relativistic CFT coupled to the SM with contact interactions and derive Sommerfeld enhancement of non-relativistic cross-sections from the theory of fermions at unitarity.
\end{quote}

\end{center}

\vfill
\noindent\line(1,0){188}
{\scriptsize{ \\ E-mail:\texttt{  \href{mailto:michele.redi@fi.infn.it}{michele.redi@fi.infn.it}, \href{andrea.tesi@fi.infn.it}{andrea.tesi@fi.infn.it}}}}
\newpage
\tableofcontents

\section{Introduction}

Two of the main paradigms for Dark Matter (DM) production are through freeze-in and freeze-out from the Standard Model (SM) thermal bath.
In the first case DM never reaches thermal equilibrium with the SM and is populated from the SM thermal bath through tiny couplings \cite{Hall:2009bx}, see 
\cite{Bernal:2017kxu} for a review. For freeze-out the situation is reversed and DM is initially in thermal equilibrium decoupling while DM becomes  non-relativistic.
In this work we extend these framework to general dark sectors (DS) showing how the physics is in all cases captured by conformal dynamics.

For freeze-in DM production takes place while SM and DM particles are relativistic so that the system can be described through a relativistic conformal field theory (CFT).
This description applies both to weakly coupled theories of elementary particles as well as strongly interacting systems.
For freeze-out of DM its also true the DM particles have no scale being strongly non-relativistic. 
In this case the initial state can indeed be described as non-relativistic  CFT. 

In this paper we put into practice the previous observation and we show how the relevant quantities that determine the DS production
can be directly extracted from CFT data, simplifying significantly the computations, in particular for multi-particle processes, see \cite{Grzadkowski:2008xi,Hong:2019nwd,Redi:2020ffc,Contino:2020tix} for related work. 
We will show that the production through freeze-in is completely determined by 2-point functions of the CFT whose structure is fixed 
up to the overall coefficient. This can be applied to weakly coupled models or to strongly coupled 
DS where no lagrangian description exists. This captures in particular 5D Randall-Sundrum models \cite{Randall:1999vf} as well as unparticles scenarios \cite{Georgi:2007ek}.
For strongly coupled theories the population of the DS depends crucially on the CFT interactions that are responsible for thermalization. 
These are in principle related to n-point functions of the CFTs. We will argue that under broad assumptions thermalization 
is instantaneous allowing to suppress the decay of dark CFT states to SM. With these ingredients we study scenarios where DM is 
an accidentally stable composite state of a spontaneously broken CFT \cite{Hong:2019nwd} analogous to glueball DM studied in \cite{Cline:2013zca,Boddy:2014yra,Soni:2016gzf,Forestell:2017wov,Acharya:2017szw,Jo:2020ggs,Redi:2020ffc,Gross:2020zam}.

CFT techniques turn out to be also relevant for freeze-out. In this case DM is non-relativistic and the relevant CFTs are the less familiar 
non-relativistic CFT that have been recently discussed in the context of unitary Fermi gas, see \cite{Nishida:2010tm} for a review.
After presenting general formulae for $n\to m$ processes we will show that Sommerfeld enhancement can be derived from the interacting CFT 
of fermions at unitarity.

The organization of this paper is as follows. In section \ref{sec:freeze-in} we lay out the general formulas for freeze-in production. 
In section \ref{sec:CFTrates} we compute the relevant rates in terms of CFT data for contact interactions. We apply these results in section
\ref{sec:applications} to weakly  and strongly coupled dark sectors. As example we consider the production from the quark portal showing
how n-particles initial states can be computed just in terms of coefficients of 2-point functions. In section \ref{sec:freezeout} we extend the discussion
to freeze-out showing that this can be related to non-relativistic CFTs. We summarize in \ref{sec:conclusions}
In  appendix \ref{appA} we extend the CFT methods to the distribution function.

\section{Freeze-in production from CFT}
\label{sec:freeze-in}

We consider DS that are singlets under SM gauge interactions. 
At sufficiently high temperatures they can be approximated by a CFT.
The simplest way (apart from gravity) to put these sectors into contact with the SM is to deform the CFT by adding contact interactions with singlet operators made of SM fields. 
The model is then simply described by,
\be\label{eq:model}
\mathscr{L}_{\rm SM}+ \mathrm{CFT} +  \sum \frac{\op_{\rm SM} \op_D}{\Lambda^{\Delta}} \,,~~~~~~~~\Delta=d_{\rm SM} + d_{\rm D} -4\,.
\ee
Here $\op_D$ is an operator of the CFT with anomalous dimension $d_D$, while $\op_{\rm SM} $ is a SM singlet operator with dimension $d_{\rm SM}$ given by the sum of
the constituents. Focusing on scalar operators, this choice selects, up to dimension 4, the following structures
\be\label{eq:SM}
\op_{\rm SM} = \{ \bar{f}_L H f_R\,,|H|^2\,,i\bar{f} \slashed{D} f\,,|D_\mu H|^2\,,-\frac 1 4 F^2, F \tilde F\}\,.
\ee
As we will see the production is entirely determined by the 2-point functions of the SM and DS operator that by conformal invariance are given by,
\be\label{eq:scalar2pt}
\langle {\rm T} \op(x)\op(0)\rangle =\frac{a_\op}{8\pi^4}\frac{1}{(x^2+i \epsilon)^d}\,.
\ee

For applications to DM we assume that the DS develops has a mass scale $M$. 
At this scale conformal invariance is completely broken and a mass gap $\sim M$ is generated. $M$ could either be dynamically generated 
or simply a bare mass term for weakly coupled theories. The SM as well can be treated as a (free) CFT with $M\sim 100$ GeV.
This is a good approximation because the DS production is dominated when the SM particles are relativistic. 
Within this assumption the DS abundance is  determined by the two-point function of the SM and CFT operators.

It will be useful to think about the production of CFT states from the SM CFT. 
In this language the SM multi-particle initial states is simply a shell of a free CFT that transition to a DS CFT state.
Naively the numerical abundance of CFT shells is controlled by the Boltzmann equation\footnote{We neglect here inverse processes. This approximation is valid as long
as the DS does not thermalize with the SM. This condition corresponds to $\gamma/T^3 < H$ that implies $T_R< \Lambda (\Lambda/\Mpl)^{1/(2\Delta-1)}$.}
\begin{equation}
\frac{dY_{\rm shell}}{dT}=- \frac{\gamma_0}{s H T}\,,
\end{equation}
where $\gamma_0$ is the spacetime density of interactions, $H$ the Hubble rate, $s=(2\pi^2/45) g_*^s T^3$ the SM entropy density, and $Y_{\rm shell}$ the number density of the CFT shell divided by the SM entropy.
By dimensional analysis
\be\label{eq:nrate}
\gamma_0 = \kappa_0\, T^4\, \bigg(\frac{T}{\Lambda}\bigg)^{2\Delta} \,,
\ee
where $\kappa_0\propto a_{\rm SM} a_\op$  will be computed precisely in the next section. Assuming that the energy density is initially negligible 
the asymptotic solution is just\footnote{We assume that the production is UV dominated so that  $T_R$ is the highest temperature of the thermal bath. This corresponds to
$2 \Delta-1>0$, i.e. $d_{\rm SM}+d_D>9/2$. For consistency $T_R\lesssim \Lambda$.}
\begin{equation}
Y_{\rm shell}\approx \kappa_0\frac {135\sqrt{10}}{2g_*^{3/2} \pi^3 (2\Delta-1)} \frac {\Mpl}{T_R} \left(\frac {T_{R}}{\Lambda}\right)^{2\Delta} \,.
\label{eq:Yshell}
\end{equation}

For free final states the number of particles is just the number of shells times the particles in each shell.
As we will discuss this quantity however is not conserved if the DS has strong interactions.
Of more general relevance is the energy density flowing into the DS. Assuming that the DS evolves relativistically the energy density is controlled by
\begin{equation}
\dot{\rho}_{\rm CFT}+ 4 H \rho_{\rm CFT} = \gamma_1 \,.
\end{equation}
The energy injection rate is now fixed to be
\begin{equation}
\gamma_1=\kappa_1\, T^5\, \bigg(\frac{T}{\Lambda}\bigg)^{2\Delta}\,,
\label{eq:rhorate}
\end{equation}
so that
\begin{equation}
\rho_{\rm CFT}(T)\approx \kappa_1\frac {3\sqrt{10}}{\sqrt{g_*} \pi (2\Delta-1)} \frac {\Mpl}{T_R} \left(\frac {T_{R}}{\Lambda}\right)^{2\Delta} T^4\,.
\label{eq:rhoab}
\end{equation}
As we will show in the next section
\begin{equation}
\kappa_{0,1}\sim \frac{a_{\op_i} a_{\op_f}}{16\pi^5}\,.
\end{equation}

A couple of remarks are in order. 
It will be important in the following that the CFT are Lorentzian. The CFT operators acting on the vacuum create momentum states of any centre of mass energy $\sqrt{s}$,
\begin{equation}
|s\,,p_0\rangle={\cal O}(p_0,\vec{p})|0\rangle\,,\quad\quad s\equiv p_0^2-|\vec{p}\,|^2>0
\end{equation}
Each shell  would thus behave as a massive particle of mass $\sqrt{s}$. Because of the connection with a continuum distribution of massive particles
the CFT shell is also known as an `unparticle' \cite{Georgi:2007ek}. As we will discuss in section \ref{sec:applications}
in interacting theories heavy CFT shells are unstable and decay to  states of lower invariant mass
quickly thermalizing the system. Viceversa if  the CFT is a weakly coupled field theory of massless particles 
the CFT shells are not coherent and the evolution  after production is determined by the individual relativistic particles. 
Thus in practise the evolution is robustly relativistic in most cases.

\begin{figure}[t]
\begin{center}
	\begin{tikzpicture}[line width=1.5 pt, scale=1]
	\draw[] (-1,1)--(0,0);
	\draw[] (-1,-1)--(0,0);
	\draw[color=blue] (0,0)--(1.5,0.5);
	\draw[dashed,color=blue] (0,0)--(1.5,0.25);
	\draw[dashed,color=blue] (0,0)--(1.5,0);
	\draw[dashed,color=blue] (0,0)--(1.5,-0.25);
	\draw[color=blue] (0,0)--(1.5,-0.5);
	\node at (2,0) {CFT};
	\draw[fill,color=gray] (0,0) circle (.4cm);
	\end{tikzpicture}\quad\quad
	\begin{tikzpicture}[line width=1.5 pt, scale=1]
	\draw[] (-1,1)--(0,0);
	\draw[] (-1,0)--(0,0);
	\draw[] (-1,-1)--(0,0);
	\draw[color=blue] (0,0)--(1.5,0.5);
	\draw[dashed,color=blue] (0,0)--(1.5,0.25);
	\draw[dashed,color=blue] (0,0)--(1.5,0);
	\draw[dashed,color=blue] (0,0)--(1.5,-0.25);
	\draw[color=blue] (0,0)--(1.5,-0.5);
	\node at (2,0) {CFT};
	\draw[fill,color=gray] (0,0) circle (.4cm);
	\end{tikzpicture}\quad\quad
	\begin{tikzpicture}[line width=1.5 pt, scale=1]
	\draw[] (-1,1)--(0,0);
	\draw[] (-1,-1)--(0,0);
	\draw[color=blue] (0,0)--(1.3,1);
	\draw[dashed,color=blue] (0,0)--(1.5,.75);
	\draw[dashed,color=blue] (0,0)--(1.5,.5);
	\draw[dashed,color=blue] (0,0)--(1.5,0.25);
	\draw[color=blue] (0,0)--(1.5,00);
	\draw[] (0,0)--(1.5,-0.75);
	\node at (2,0.5) {CFT};
	\draw[fill,color=gray] (0,0) circle (.4cm);
	\end{tikzpicture}\quad
	\begin{tikzpicture}[line width=1.5 pt, scale=1]
	\draw[] (-1,1)--(0,0);
	\draw[] (-1,0)--(0,0);
	\draw[] (-1,-1)--(0,0);
	\draw[color=blue] (0,0)--(1.3,1);
	\draw[dashed,color=blue] (0,0)--(1.5,.75);
	\draw[dashed,color=blue] (0,0)--(1.5,.5);
	\draw[dashed,color=blue] (0,0)--(1.5,0.25);
	\draw[color=blue] (0,0)--(1.5,00);
	\draw[] (0,0)--(1.5,-0.75);
	\node at (2,0.5) {CFT};
	\draw[fill,color=gray] (0,0) circle (.4cm);
	\end{tikzpicture}

\end{center}
\caption{\label{fig:freeze-in} \it Relativistic freeze-in. A list of processes relevant for the population of the DS CFT from SM initial states (black lines). Notice that we also include reactions where a SM field is present in the final states and/or more than two SM particles are present in the initial state.}
\end{figure}
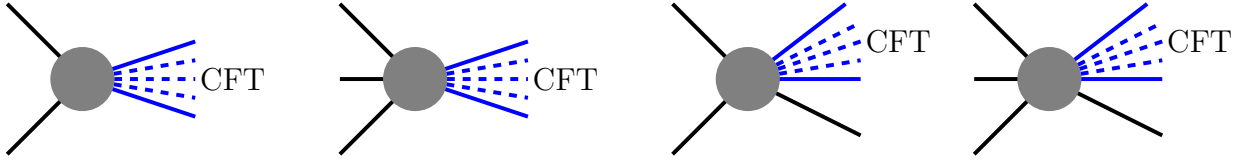

It is worth emphasising that a contact interaction can mediate more than one process. For example the operator $\bar{f}_L H f_R$ allows to produce the DS from 3 SM particles or from 2 SM particles in association with a SM, see  figure \ref{fig:freeze-in}). In the latter case, thanks to the approximate conformal symmetry of both the dark and visible sector, we can think of the final state as excited by a composite operator build up with $\op$ and SM fields. All these contributions are of the same order so they must be added to obtain the total production.

\section{CFT rates}
\label{sec:CFTrates}

The central objects of our discussion are the generalized space-time density rates $\gamma_i$ in eq.s (\ref{eq:nrate}) and (\ref{eq:rhorate}). 
These are defined as\footnote{We neglect for simplicity quantum statistics. For free CFTs such as the SM this introduces an error at most of order 10\%, see appendix for details.}
\be\label{eq:gamma}
\gamma_n=\int d\Pi_i d\Pi_f \, p_0^n\, e^{-p_0/T}|\mathcal{M}_{i \to f}|^2 (2\pi)^4 \delta^{(4)}(p_i-p_f) \,,
\ee
where $d\Pi_{i,f}$ are the phase space densities for initial and final states. Rather than considering initial and final states as made of particles it will be useful to think of them as energy momentum shells.
The transition amplitude between initial and final shells is just,
\begin{equation}
 {\cal M}_{i\to f}=\frac 1 {\Lambda^{\Delta} } \langle p_f |  \op_f(0) \op_i(0) |p_i\rangle=\frac 1 {\Lambda^{\Delta} } \langle p_f |  \op_f(0)|0\rangle\langle0| \op_i(0) |p_i\rangle\,.
\end{equation}
This formula applies to any type of contact interactions as in eq.~\eqref{eq:model}. In the above equation, $|p\rangle\propto |s, \vec{p}\rangle$ are an orthonormal complete set of momentum states. \begin{equation}
\int_{p_0\ge|\vec{p}|} \frac{d^4p}{(2\pi)^4} |p\rangle \langle p|=1\,.
\label{eq:completeset}
\end{equation} 
The completeness relation allows us to write the space-time density rate as
\begin{equation}
\gamma_n=\int_{p_0\ge|\vec{p}|} \frac{d^4 p}{(2\pi)^4}  e^{-p_0/T} p_0^n\, |\langle 0| \op_i(0) |p\rangle|^2 |\langle 0| \op_f(0) |p\rangle|^2\,.
\end{equation}
By Lorentz invariance the matrix elements are functions of $s=p^2$. Performing the integral over $p_0$ one finds
\begin{equation}
\gamma_n=\frac T {8\pi^3}\frac {\partial^n}{\partial(-\beta)^n} \int ds\, \sqrt{s}\,K_1(\beta \sqrt{s})\, |\langle 0| \op_i(0) |p\rangle|^2 |\langle 0| \op_f(0) |p\rangle|^2 \,,\quad\quad\beta\equiv \frac 1 T\,,
\label{eq:gamman}
\end{equation}
where $K_1(x)$ is the Bessel function of second kind.

The matrix elements appearing in the formula above are related to the Wightman 2-point function. 
Indeed inserting the complete set of states (\ref{eq:completeset}) we obtain,
\begin{equation}
\langle O(x) O(0)\rangle = \int_{p_0\ge|\vec{p}|} \frac{d^4p}{(2\pi)^4} e^{-i p x}  |\langle 0| \op(0) |p\rangle|^2
\end{equation}
so that the matrix element is the Fourier transform of the Wightman 2-point function.
The latter in turn is related to T-ordered 2-point function by\footnote{The T-ordered 2-point function can alternatively be derived
using the optical theorem as in \cite{Redi:2020ffc,Contino:2020tix,Garani:2021zrr}. 
The use of the optical theorem allows to extend the derivation beyond contact interactions, see section \ref{sec:freezeout}.}
\begin{equation}
|\langle 0| \op(0) |p\rangle|^2 = 2 \mathrm{Im}[i \langle {\rm T}\op(p)\op(-p)\rangle]\,.
\end{equation}

So far the derivation has been general and it does not depend on the assumption of  conformal invariance. 
However, in the special case of conformal sectors the structure of the 2 point functions is fixed up to the coefficient $a_{\op}$ allowing to derive explicit formulas
to which we turn. For operators with spin analogous results can be derived.

\subsection{Scalar operators}
The CFT production rate that originates from the  contact interaction between two scalar operators can be derived as follows. The time-ordered scalar 2-point function is given in eq. (\ref{eq:scalar2pt}), and its Fourier transform then reads
\be
\quad \langle {\rm T} \op(p)\op(-p)\rangle =-i \frac{a_\op}{2\pi^2}\frac{\Gamma(2-d)}{4^{d-1}\Gamma(d)}(-p^2)^{d-2}\,.
\ee
Taking the imaginary part we are left with
\be
 \mathrm{Im}[i \langle \op(p)\op(-p)\rangle]=\frac{a_\op}{2\pi}\frac{d-1}{4^{d-1}\Gamma(d)^2}(p^2)^{d-2}\theta(p^2) \theta(p_0)\,.
\ee
This quantity is needed to perform the integral in eq. \eqref{eq:gamman}. For the case of interest, we find the coefficients in eqs. (\ref{eq:nrate}), (\ref{eq:rhorate}) to be 
\begin{eqnarray}
\kappa_0 &=& \frac{a_{\op_i} a_{\op_f}}{32\pi^5}\, \frac{\Gamma(d_i+d_f-2)\Gamma(d_i+d_f-3)}{\Gamma(d_i)\Gamma(d_i-1)\Gamma(d_f)\Gamma(d_f-1)}\,, \\
\kappa_1 &=& \frac{a_{\op_i} a_{\op_f}}{16\pi^5}\, \frac{\Gamma(d_i+d_f-1)\Gamma(d_i+d_f-3)}{\Gamma(d_i)\Gamma(d_i-1)\Gamma(d_f)\Gamma(d_f-1)}\,.
\label{eq:gammaB}
\end{eqnarray}
For a fixed combination of $d_i+d_f=\Delta+4$, the largest value of $\gamma$ is given by $d_i=d_f$. This implies that for multiparticle 
transitions the most relevant processes are the ones with similar number of particles. Note that the rates vanish for $d_{i,f}=1$ corresponding to elementary 
single particle state. In this case the process is kinematically forbidden for finite masses and by continuity is also zero in the massless limit.

\subsection{Fermionic operators}
Let us now consider the contact interactions between fermionic operators. 
The 2-point function of fermionic operators of dimension $d$ is given by
\be
\langle {\rm T}\op_\alpha(x)\op_{\dot{\beta}}(0)\rangle  =i\frac{b_\op}{8\pi^4}\frac{x_\mu \sigma_{\alpha\dot{\beta}}^\mu}{(x^2)^{d+1/2}}=\frac{b_\op}{8\pi^4}\frac 1 {2d-1} (-i\partial_\mu)
\frac {\sigma_{\alpha\dot{\beta}}^\mu}{(x^2)^{d-1/2}}\,.
\ee
The Fourier transform is
\be
\quad \langle {\rm T} \op_\alpha(p)\op_{\dot{\beta}}(-p)\rangle= -i\frac{b_\op}{2\pi^2}\frac{\Gamma(5/2-d)}{4^{d-3/2}(2d-1)\Gamma(d-1/2)} [p_\mu\Gamma_{\alpha\dot{\beta}}^\mu] (-p^2)^{d-5/2}\,.
\ee
The computation of the space-time density rate is the same as for scalars, after taking the imaginary part of the Fourier transform, we can compute the integral in \eqref{eq:gamman}, by performing explicitly the contraction of the spin indices. In this case we arrive at the expressions
\begin{eqnarray}
\kappa_0&=&\frac{b_{\op_i} b_{\op_f}}{16\pi^5}\, \frac{\Gamma(d_i+d_f-2)\Gamma(d_i+d_f-3)}{\Gamma(d_i+1/2)
\Gamma(d_i-3/2)\Gamma(d_f+1/2)\Gamma(d_f-3/2)}\,, \\
\kappa_1&=&\frac{b_{\op_i} b_{\op_f}}{8\pi^5}\, \frac{\Gamma(d_i+d_f-1)\Gamma(d_i+d_f-3)}{\Gamma(d_i+1/2)
\Gamma(d_i-3/2)\Gamma(d_f+1/2)\Gamma(d_f-3/2)}\,, 
\label{eq:gammaF}
\end{eqnarray}
that nicely vanish when $d_{i,f}=3/2$.

\subsection{Vector and tensor operators}

For completeness we also provide the relevant formulae for conserved vector, $\partial_\mu J^\mu=0$, and tensor operators, $\partial_\mu T^{\mu\nu}=0$.

\paragraph{Conserved current}~\\
A conserved current has dimension 3 and the 2-point function can be written as
\be
\langle J_\mu(x)J_\nu(0)\rangle  =\frac{c_J}{8\pi^4}\frac{1}{(x^2)^3}I_{\mu\nu}\,,~~~~~~~I_{\mu\nu}\equiv(\eta_{\mu\nu}-2 \frac{x_\mu x_\nu}{x^2})\,,
\ee
with Fourier transform
\be
\langle J_\mu(-p)J_\nu(p)\rangle  =-i\frac{c_J}{96\pi^2} p^2 \log(-p^2) P_{\mu\nu} \,,~~~~~~~~~~ P_{\mu\nu} \equiv \eta_{\mu\nu}-p_\mu p_\nu/p^2
\ee
Considering current-current  interaction in eq.\eqref{eq:model}, $J_\mu^i J^{\mu\,f}/\Lambda^2$,  we get
\be
\gamma_0=\frac{c_{J_i} c_{J_f}}{8\pi^5}\,  \left(\frac{T}{\Lambda}\right)^{4}\,T^4\,,~~~~~~~~~~\gamma_1=\frac{c_{J_i} c_{J_f}}{\pi^5}\,  \left(\frac{T}{\Lambda}\right)^{4}\,T^5\,.
\ee

\paragraph{Conserved spin-2 tensor}~\\
For conserved spin-2 tensors the two-point function reads \cite{Gubser:1997se}
\begin{equation}
\langle T_{\mu\nu}(x) T_{\rho \sigma }(0)\rangle = \frac 1{4\pi^4} P_{\mu\nu  \sigma\rho} \frac {c_T} {x^8}\,, \quad P_{\mu\nu\sigma\rho}\equiv\frac 1 2(I_{\mu\sigma} I_{\nu\rho}+I_{\mu\rho}I_{\nu\sigma})-\frac 1 4 \eta_{\mu\nu}\eta_{\sigma\rho} \,,
\end{equation}
with Fourier transform
\begin{equation}
\langle T_{\mu\nu}(p) T_{\rho \sigma }(-p)\rangle= \frac{c_T}{7680 \pi^2}  \left(2P_{\mu\nu} P_{\rho\sigma}-3P_{\mu\rho} P_{\nu\sigma}-3P_{\mu\sigma} P_{\nu\rho}\right)  \log(-p^2)\,.
\end{equation}
For dimension 8 contact interaction $T_{\mu\nu}^i T^{\mu\nu\,f}/\Lambda^4$ one finds
\be
\gamma_0=\frac{9c_{T_i} c_{T_f}}{2\pi^5}\,  \left(\frac{T}{\Lambda}\right)^{8}\,T^4\,,~~~~~~~~~~\gamma_1=\frac{54 c_{T_i} c_{T_f}}{\pi^5}\,  \left(\frac{T}{\Lambda}\right)^{8}\,T^5\,.
\ee
See also \cite{Bernal:2018qlk} for an application to freeze-in via massive spin-2 mediators.

\section{Phenomenological Applications}
\label{sec:applications}

We now outline two applications of our formalism in the context of DM at weak coupling and strong coupling. Applications to dark radiation
can also be considered. While the production of the DS is completely determined by an exact CFT, in order to discuss DM we will have to introduce breaking
of scale invariance. We will assume that this generates a mass gap. If the DS is a singlet under the SM it will only communicate
through contact interactions and gravity with the SM. This automatically leads to accidental DM candidates \cite{Antipin:2015xia} where DM cosmological stability
follows automatically from the structure of theory. See \cite{Kribs:2016cew} and \cite{Beylin:2020bsz} for reviews on dark sectors with strong interactions.

\subsection{Weakly coupled Dark Sector}

Let us first consider the production of a weakly coupled DS. An example of such realization can be a non-abelian gauge theory of gluons and fermions as discussed in \cite{Redi:2020ffc,Dondi:2019olm,Garani:2021zrr}.
At energies or temperatures larger than the confinement scale the sector is approximately a free field theory of massless particles so one could directly obtain the freeze-in production computing the cross-section with standard Feynman diagrams \cite{Garny:2017kha}.
It is much easier, however,  to extract the rates exploiting the fact that initial and final states are CFT states as this does not require complicated phase space integrals.
Importantly after production the particles that make up each shell do not propagate coherently but independently. If the operator $\mathcal{O}_D$, appearing in \eqref{eq:model}, contains $n$ fields 
each CFT shell is made of $n$ DS particles that are relativistic so that
\begin{equation}
Y_D=n\, Y_{\rm CFT} \,,~~~~~~~~~~~~~~~~\rho_D= \rho_{\rm CFT}\,.
\end{equation}
Let us note that while the  densities are simply obtained from the CFT, the determination of the phase space distribution function is less trivial.
We can compute the distribution of CFT shells introducing the appropriate collision term as explained in appendix \ref{appA}. 
Because however the CFT states describe a continuum of massive particles their evolution is not trivial.
One can think of the final particles as arising from the instantaneous decay of the CFT shells and derive the effective collision term 
from the CFT one, see appendix.

The loss of coherence has two important effects. First, since the particles are massless the energy density dilutes like radiation, $\rho \propto 1/a^4$. Second, each particle is stable and the annihilation rate to the SM is very suppressed by the low density of the DS. 

Depending on the size of interactions two things can happen:

If the interactions are weak as the universe expands the mass gap of the DS (either dynamically generated or the mass of the particles)
becomes important and the energy density is converted into non-relativistic particles that make up DM. 
In this case the abundance of DM is simply determined by $Y_{D}$. Requiring that the correct DM density is reproduced
$Y_{D} M\sim $ eV we find a relation between the DM, reheating temperature and the scale suppressing the higher dimensional 
operators,
\begin{equation}
\frac {\Omega h^2}{0.12}=\frac {Y_{D} M_{\rm DM}}{0.4\,{\rm eV}}\approx 1.6\times 10^7\,\frac{M_{\rm DM}}{\rm GeV} \frac{\kappa_0}{2\Delta -1} \frac{\Mpl}{\Lambda}\left(\frac {T_R}{\Lambda}\right)^{2\Delta-1}\,.
\label{eq:abundancefree}
\end{equation}
For $\Delta=2$ (dimension 6 operator) and $\Lambda=\Mpl$ this reproduces the estimate for gravitational production through graviton exchange \cite{Redi:2020ffc}.

If the interactions become sizable as in the presence of (dark) non-abelian gauge interaction the DS will thermalize in the relativistic regime. 
Using conservation of energy, $\rho_{\rm CFT}\sim g_D T_D^4$, this leads to a DS temperature $T_D$ defined by the following ratio with the SM temperature
\begin{equation}
\xi\equiv \frac {T_D}{T}\approx \frac{1}{(2\Delta-1)^{1/4}} \left(\frac {\kappa_1}{g_D}\right)^{1/4} \left(\frac {\Mpl}{T_R}\right)^{1/4}  \left(\frac {T_R}{\Lambda}\right)^{\Delta/2}\,,
\label{eq:ratioT}
\end{equation}
where $g_D$ is number of degrees of freedom of the DS. Note that the low density generates a small temperature of the DS
and that thermalization increases the number of particles. From here the DS begins its own thermal history, see \cite{Garani:2021zrr}
for the discussion in the context of dark QCD.
If the sector for example confines at a critical temperature $T_c\sim f$ the energy of the thermal bath is converted into DM particles.
Assuming that they are non-relativistic as for glueballs we can estimate the  abundance as
\begin{equation}
\frac {\Omega h^2}{0.12}=\frac{\rho_{\rm CFT}(T_c)}{0.4\,{\rm eV}\,s(T_c/\xi)} \approx 1.7\times 10^7\, \frac{1}{(2\Delta-1)^{3/4}}\, \frac{f}{\rm GeV}\left(\frac{\kappa_1}{g_D}\right)^{3/4}\left(\frac{\Mpl}{\Lambda}\right)^{3/4}\left(\frac {T_R}{\Lambda}\right)^{\frac 3 2\Delta-\frac 3 4}\,.
\label{eq:abundanceth}
\end{equation}
We might also consider the possibility that the DM particles are much lighter than $T_c$. 
This is realized if the DM is a Nambu-Goldstone boson from the spontaneous breaking of global symmetries.
In this case the DM particles are relativistic at production. Assuming that the system thermalizes after the phase transition the DM abundance is given
by (\ref{eq:abundanceth}) with $f$ replaced by $M_{\rm DM}$.

\subsection{Coherent Dark Sector}

If the DS is strongly interacting the physics  is quite different from the one discussed above.
In this case the CFT shells are coherent states that behave as single particle states. Depending 
on the nature of the DS interactions different scenarios are realized.

\paragraph{Generalized free fields:}

In the literature ``free'' CFTs with trivial n-point function are often considered, see for example \cite{Dymarsky:2014zja}. Such theories 
can be described by a non-local action that formally satisfy the axioms of CFT even though it is unclear whether they 
are fully consistent due  to the lack of an energy momentum tensor. In this case we can attach a precise meaning to the 
CFT shells since their number is conserved in the DS.

By integrating the Boltzmann equations (see appendix) one finds
\be
n_{\rm shell} \sim T^3 \,\frac{\Mpl}{T_R}\left( \frac{T_R}{\Lambda}\right)^{2\Delta}\,, \quad \rho_{\rm CFT} \sim T^3 T_R\, \frac{\Mpl}{T_R}\left( \frac{T_R}{\Lambda}\right)^{2\Delta}\,,\, p_{\rm CFT} \sim\frac{ T^5}{ T_R}\, \frac{\Mpl}{T_R}\left( \frac{T_R}{\Lambda}\right)^{2\Delta}
\ee 

This result however does not take into account that each shell can decay through the same process that produces them. We estimate the decay rate $\Gamma$ as
\begin{equation}
\frac {d\Gamma(s,p_0)}{ds} \sim \frac 1{p_0}  \left( \frac s {\Lambda^2}\right)^{\Delta}\,.
\end{equation}
Taking this into account the abundance would be actually negligible since the abundance is dominated by heavy CFT shells that immediately decay back
to SM. Therefore, in order to realize a viable DM scenario one should include a stabilizing symmetry that forbids direct decays (and production) to the SM,
see \cite{Kikuchi:2007az,Csaki:2021gfm}.

\paragraph{Strongly coupled:}
If the interaction in the DS are strong the non-thermal CFT states will decay to lower mass shells.
By conformal invariance the rate scales as $\Gamma_{dec}\propto T$ so it is much faster than the production rate unless the effective coupling is exceedingly small. 
The fast degradation of energy allows the DS to survive until today as the lighter CFT shells are more stable.

The endpoint of the energy degradation is thermal equilibrium where
\begin{equation}
\rho=3p = g_D\frac {\pi^2}{30} T_D^4\,.
\end{equation}
If the thermalization process is instantaneous we can simply obtain the DS temperature using conservation of energy so that eq. (\ref{eq:ratioT})
applies. In this regime the full Boltzmann equation for the density has the form
\begin{equation}
\dot{\rho}_{\rm CFT}+ 4 H \rho_{\rm CFT} = \gamma_1(T)-\gamma_1(T_D)\,.
\end{equation}
Since $T_D\ll T$ the energy loss from the thermal bath is negligible.

Thermalization of strongly coupled CFT was studied through holography \cite{Bhattacharyya:2009uu,Wu:2013qi} with the result that thermalization completes in a time $1/T_D$. 
Given that the production is dominated by the highest temperature this implies
\begin{equation}
\tau_{\rm therm}\sim \frac 1 {T_R}\left(\frac {T_R}{\Mpl}\right)^{1/4}  \left(\frac {\Lambda}{T_R}\right)^{\Delta/2}\,.
\end{equation}
Since this is always smaller than $H$ it follows that thermalization is practically instantenous and the system evolves relativistically.

The final result is that despite very different physical origin the DS is populated similarly to the weakly coupled scenario. The DM abundance is
again given parametrically by the estimates of eq.s (\ref{eq:abundancefree}) and (\ref{eq:abundanceth}).

\subsection{Example: Conformal DM}

\begin{figure}[t]
\centering
\includegraphics[width=0.8\linewidth]{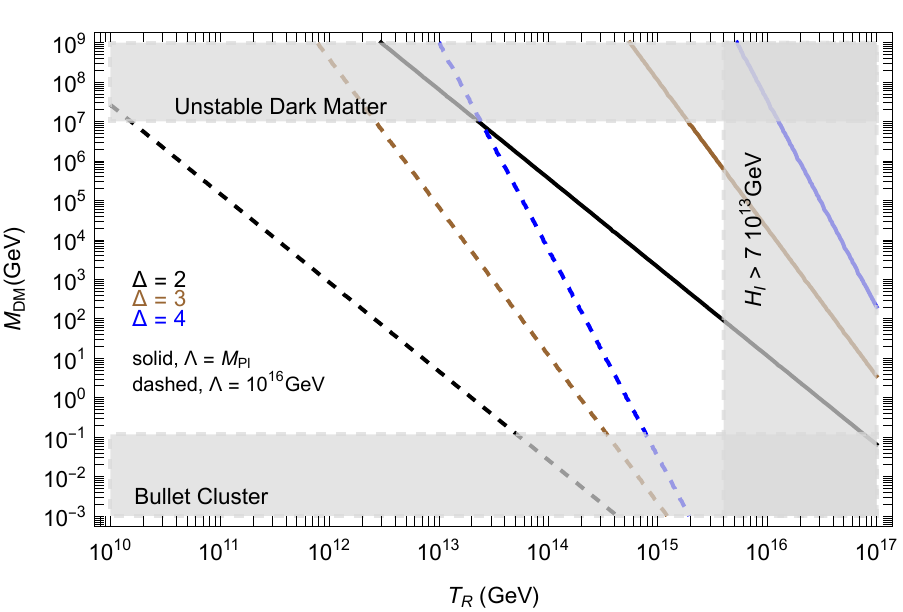}
\caption{\label{fig:abundance}\it Phase diagram of conformal DM produced through a Yukawa coupling for dimension 2, 3 and 4of the dark sector operators.
We assume the contact interaction to be suppressed by $\Lambda=\Mpl=2.4\cdot 10^{18}$ GeV (solid lines) and $\Lambda=10^{16}$ GeV (dashed lines), $g_D=1$ and $M_{\rm DM}=5 f$. 
The grey regions are excluded respectively by DM self-interaction taking $\sigma_{\rm el}= \pi/M_{\rm DM}^2$, scale of inflation and DM lifetime 
where we use the estimate $\tau_{\rm DM}\sim \Mpl^4/M_{\rm DM}^5$.}
\label{fig:abundannce}
\end{figure}

We now consider the production of a conformal DS through scalar operators.
The anomalous dimension of SM operators are simply obtained from the elementary 2-point functions
\begin{equation}
\langle {\rm T} \phi(x) \phi(0)\rangle = \frac 1 {4\pi^2}\frac 1 {x^2+i \epsilon}\,,\quad \langle {\rm T} \psi_{\alpha}(x) \bar{\psi}_{\dot{\beta}}(0)\rangle = \frac i {2\pi^2}\frac {x_\mu \sigma_{\alpha\dot{\beta}}^\mu} {x^4+i \epsilon}\,.
\end{equation}
The SM composite operators have thus
\begin{equation}
a_{|H|^2}=4\,,~~~~~~a_{i\bar{f}\slashed{\partial} f}=0\,,~~~~~~~~a_{\bar{f}_L^i H^i f_R+h.c.}=\frac 4 {\pi^2}\,,~~~~~~a_{|\partial H|^2}=48\,,~~~~~~a_{F^2/4}=24\,.
\end{equation}

With these ingredients, using the formulae (\ref{eq:gammaB}), (\ref{eq:gammaF}) the energy density of the DS is given by eq. (\ref{eq:rhoab}) with the following 
coefficients
\begin{center}
\begin{tabular}{c||c|c|c|c}
 &  $16\pi^5\kappa_1(d_D=2)/a_{\rm D}$ & $16\pi^5\kappa_1(d_D=3)/a_{\rm D}$ & $16\pi^5\kappa_1(d_D=4)/a_{\rm D}$ \\ 
\hline 
$|H|^2\to CFT$ & $8$  & $12$ & $16$  \\
$|D_\mu H|^2 \to CFT$  & 192  & 1440 & 5760 &  \\
$F^2/4 \to CFT$  & 96 & 720 & 2880 &  \\
\hline 
$\bar{f}_L H f_R\to CFT $ & $16/\pi^2$   & $120/\pi^2$ & $480/\pi^2$    \\
$f_{L,R} H\to CFT +f_{R,L}$  &  $64/\pi^2 \times 1/2$  & $120/\pi^2\times 1/2$ & $192/\pi^2\times 1/2$  &   \\
$f_L f_R \to CFT+ H$  &  $48/\pi^2\times 1/2$  & $120/\pi^2\times 1/2$ & $240/\pi^2\times 1/2$  
\end{tabular}
\end{center}

The result is particularly non-trivial for the Yukawa portal, the last three rows of the above table.
Since the SM operator contains 3 SM particles this induces $3\to$ CFT and $2\to \rm{CFT}+1$ production channels.  
Both processes are easily captured with our formalism with no need to study complicated phase space integrals.
For  the first process we can directly use the formula in the previous section.
For the production in association $f H \to {\rm CFT} + f$ or $\bar{f} f\to {\rm CFT}+ H$ we can effectively use final state
operators $O_{\rm CFT}X$. Note that in this case $b_{CFT X}= a_{CFT} a_{X}$ so that $b_{\op_i} b_{\op_f}= a_{\op_i} a_{\op_f}$.
Single production and associated production contribute parametrically the same amount to the production, we include a factor of $1/2$ for the associated production to take into account that about half of the energy flows into the SM particle produced in association.

Assuming that the DS develops a mass gap $f$ the lightest state is accidentally stable and gives rise to 
a good DM candidate. Explicit examples are for example glueballs and pions in dark gauge theories studied in \cite{Redi:2020ffc,Garani:2021zrr}. 
Assuming the lightest state to be excited by ${\cal O}_D$ we estimate the decay rate into the SM,
\begin{equation}
\Gamma_{\rm SM} \sim M_{\rm DM} \left(\frac {M_{\rm DM}}{\Lambda}\right)^{2\Delta}\left( \frac f {M_{\rm DM}}\right)^{2d_D-2}\,.
\end{equation}
Bosonic  DM may also decay into gravitons with a rate $\Gamma_{\rm GR}\sim M_{\rm DM}^5/\Mpl^4$.

In figure \ref{fig:abundance} we show the parameter space of conformal DM. We plot the mass of DM as function of the reheating temperature in order to reproduce the cosmological 
abundance. In the figure we assume Planck suppressed quark portal couplings, $\Lambda=\Mpl$ in the figure.

\section{Freeze-out and Non Relativistic CFTs}
\label{sec:freezeout}

For freeze-out we are interested in the annihilation of non-relativistic DM particles into particles of the thermal bath.
This initial state can be interpreted as a state of a non-relativistic CFT \cite{Nishida:2007pj} allowing to generalize the previous discussion to freeze-out.
The  CFT approach gives a different derivation of the thermally averaged annihilation cross-section \cite{Gondolo:1990dk} that immediately generalizes 
to N-particle initial states. Moreover we will see that Sommerfeld enhancement can be derived from an interacting NR CFT in the univerality class of
fermions at unitarity \cite{Nishida:2010tm}.

We will be interested in general annihilation processes so the derivation will be slightly different than in section \ref{sec:CFTrates} 
where only contact interactions were considered. Using the optical theorem we write the space-time density of interactions (\ref{eq:gamma}) as
\be
\gamma_0= \int d\Pi_i e^{-p_0/T} 2\mathrm{Im}(\mathcal{M}^{\rm forward}_{i \to f^* \to i})\,.
\ee
As in the relativistic unparticle case the phase space of non-relativistic initial particles is determined by conformal invariance, see \cite{Hammer:2021zxb}
for a recent discussion in the context of nuclear physics. In order to derive an explicit formula for the  (imaginary part of the) amplitude more details about the theory have to be known. If the interaction can be parametrized in the form of effective operators as in \cite{Hammer:2021zxb}, suppressed by some UV scale, the discussion would be in principle similar to the previous section.

The 2-point function of scalar operators of NR CFTs is fixed to be,
\begin{equation}
\langle T \psi(t,x)\psi^*(0,0)\rangle=i C \theta(t)\left(\frac 1 {i t}\right)^{\Delta}e^{-M x^2/(2 t)}\,,
\end{equation}
whose Fourier transform reads,
\begin{equation}
\langle {\rm T}O(\omega,p) O(-\omega,-p)\rangle=- i C \left(\frac {2\pi}{M}\right)^{3/2} \Gamma(5/2-\Delta) \left(\frac {p^2}{2M}-\omega \right)^{\Delta - 5/2}\,.
\end{equation}
Note that differently from relativistic CFTs the 2-point function depends both on the scaling dimension and on 
the conserved number of particles through the total mass $M=N m$. For $\Delta=3/2$ and $N=1$ we recover the free field limit
with $C=(M/2\pi)^{3/2}$.

As in the relativistic case the phase space of N non-relativistic particles is related to the 2-point function of the operator,
\begin{equation}
d\Pi_{\rm NR}= \rm{Im}[i \langle T\op(\omega,p)\op(-\omega,-p)\rangle] \frac {d\omega d^3p}{(2\pi)^4} \,.
\end{equation}
A generic CFT state with $\Delta \ne 3/2N$ has phase space as a fractional number of particles and because of this has been named an unparticle.

The formula above allows us to rewrite the space-time density rate as
\begin{equation}
\gamma_0= \frac C {(2m)^N}\left(\frac {2\pi}{M}\right)^{3/2} \frac {2\pi}{\Gamma(\Delta-3/2)} \int \frac{d^4p}{(2\pi)^4} e^{-p_0/T} \left(\omega - \frac {p^2}{2M}\right)^{\Delta - 5/2} \theta\left(\omega - \frac {p^2}{2M}\right)  \mathrm{Im}(\mathcal{M}^{\rm forward}_{i \to f^* \to i})
\end{equation}
where we have supplied the factor $(2m)^N$ to agree with the relativistic normalization of the amplitude. 
Using $p_0=M+\omega$ and the change of variables $s=p_0^2-|\vec{p}\,|^2$ we obtain
\be
\begin{split}
\gamma_0&=\frac 1 {4\pi^3}\frac C {(2m)^N} \left(\frac {2\pi}{M}\right)^{3/2} \frac {1}{\Gamma(\Delta-3/2)} 
\times\\
& \times \int_{M^2}^\infty ds \mathrm{Im}(\mathcal{M}^{\rm forward}_{i \to f^* \to i})\int_{\sqrt{s}}^\infty dp_0  e^{-p_0/T} \sqrt{p_0^2-s} \left(\frac{s-M^2-\omega^2}{2M}\right)^{\Delta - 5/2}\,.
\end{split} 
\ee
In the NR limit we neglect $\omega^2$ obtaining,
\begin{equation}
\gamma_0=\frac C {(2\pi)^N}T \frac {2^{2-\Delta}} {\sqrt{\pi}\Gamma[\Delta-3/2]} M^{1-\Delta} \int_{M^2}^\infty ds\,\mathrm{Im}(\mathcal{M}^{\rm forward}_{i \to f^* \to i})\,   \left(s-M^2\right)^{\Delta-5/2} \sqrt{s} K_1\left[ \frac{\sqrt{s}}T\right] 
\label{eq:gammaNR}
\end{equation}

\paragraph{Free initial state:}
If the initial state is just free particles we have\footnote{The normalization can be simply obtained from the 2-point function of
a free non relativistic particle $\langle T \phi(t,x)\phi(0,0)\rangle=\theta(t)(\frac m {2\pi i t})^{\frac 3 2}e^{-m x^2/(2 t)}$ using the fact that the N particle state is the product of N free fields.},
\begin{equation}
\Delta = 3/2 N\,,~~~~~~~~~M =N m\,,~~~~~~~~~~~C=\left(\frac {m}{2\pi}\right)^{\frac 3 2 N}
\end{equation}
For the special case $N=2$ using $\sigma(s)\sqrt{s(s-4m^2)}=\mathrm{Im}(\mathcal{M}^{\rm forward}_{i \to f^* \to i})$
\begin{equation}
\gamma_0=\frac 1 {2m}\frac{T}{64\pi^2}  \int_{4m^2}^\infty ds\, \sigma(s) (s-4m^2) s   K_1\left[\frac{\sqrt{s}}T\right] 
\end{equation}
which reproduces the famous formula for the thermally averaged cross-section \cite{Gondolo:1990dk} in the NR limit where $s\approx 4m^2$. 
This immediately generalizes to n-particle initial states.

\paragraph{Sommerfeld enhancement:}

Remarkably eq. (\ref{eq:gammaNR}) can be directly applied to interacting particles in the initial state. In particular 
if the annihilating non-relativistic particles have a large scattering length they are described by an interacting 
CFT in the same universality class of fermions at unitarity \cite{Nishida:2007pj}. 

For this theory it is known that the anomalous dimension of the operator with $N=2$ is exactly $\Delta=2$ rather than $\Delta=3$ for free particles.
Therefore effectively the cross-section is thus multiplied by the factor,
\begin{equation}
S_2= \frac {A_2}{s-4m^2}=  \frac{A_2}{p^2}\,.
\end{equation}
This situation is precisely realized when the particles in the initial state interact with a short range force. 
In correspondence of the values of the mass where the potential supports zero energy bound states 
the s-wave cross-section is enhanced by the factor,
\begin{equation}
{\rm SE}\sim \frac {a_0^2/r_0^2}{1+p^2 a_0^2}\,,
\end{equation}
where $a_0$ is the scattering length and $r_0$ the effective range. This is known in the DM literature as Sommerfeld enhancement.
In our approach we are expanding the theory around the strongly interacting fixed point $a_0=\infty$.
We can think of the long distance interactions as preparing a CFT state that then annihilate through short distance interactions. 
Note that the proportionality factor $A_2$ cannot be fixed from first principles.  
This corresponds to the fact that  the height of the Sommerfeld peaks is determined by the UV physics such us the depth of the potential. 

Interestingly we can extend this to initial states with 3 particles relevant for $3\to n $ processes. The very same non-relativistic CFT predicts 
$\Delta =4.27272$ \cite{Tan,Werner:2006zz} for the lowest operator with $N=3$.  This allows to conjecture the generalized Sommerfeld enhancement factor for 3 fermionic particles as
\begin{equation}
S_3=\frac {A_3}{(s-9m^2)^{0.23}}\,.
\end{equation}
We leave a detailed study to future work.

\section{Summary}
\label{sec:conclusions}

In this work we introduced a new framework to study freeze-in and freeze-out of DM.
The key observation is that in both cases the initial multi-particle states can be approximately described as CFT states,
respectively relativistic or non-relativistic,  labeled by their total momentum $\vec{p}$ and center of mass energy $\sqrt{s}$. This allows to determine 
the relevant rates in terms of CFT data with no need of complicated phase  space integrals.

In the case of perturbative DS the results reproduce and generalize textbook results.
For example we can easily compute cross-sections and rates corresponding to $n\to m$ processes through the 2-point functions of CFT operators. 
When the DS is strongly coupled our technique becomes necessary and provides a generalization of Boltzmann's equations to CFTs.

From a physical point of view weakly and strongly coupled DS behave quite differently.
In the first case, while the production is fully determined by CFT multiparticle states, the subsequent evolution is described by free massless particles.
This implies that the DS evolves as radiation and that the DS particles are accidentally stable.
In the strongly coupled regime the CFT states are coherent due to strong interactions and behave as single particles after production. 
This implies that the CFT states can decay through the same process that allows for their production.
If this were the end of the story the abundance of the DS would be negligible. Including interactions
however allows the heavy CFT shells to rapidly decay to lower mass shells rapidly reaching thermalization. 
Because the DS ends up being much colder than the visible sector this leads to negligible decay into the SM and to an energy density comparable
to weakly coupled models also behaving as radiation. With these ingredients we studied in general the freeze-in production of conformal
DS from the SM thermal bath through contact interactions. Assuming that this sector has a mass gap this produces
DM scenarios where the mass is determined by the reheating temperature and dimensionality of the portal.

The CFT approach can also be generalized to freeze-out. 
Here the initial state is non-relativistic and can be interpreted as a state of recently discussed non-relativistic CFTs.
In the simplest case the CFT is just a free non-relativistic field theory described by the Schr\"oedinger action.
Interestingly Sommerfeld enhancement can be understood in terms of an interacting CFT describing ``fermions at unitarity''.
This also allows to generalize Sommerfeld enhancement to $N$ particle annihilation processes.

This work opens the way to a more systematic use of CFT techniques to study relativistic and non-relativistic plasmas.
The dynamics of conformal DS raises several questions. In the case of interacting CFTs self-interactions play a crucial role for
the  viability of the scenario. These are in principle determined by CFT n-point functions. It would be interesting 
in this context to study the decays and thermalization directly from CFT data using the generalized S-matrix observables of \cite{Gillioz:2018mto,Gillioz:2020mdd}. 
For non-relativistic CFT the connection between fermions at unitarity and Sommerfeld enhancement deserves further study.

{\small
\subsubsection*{Acknowledgements}
MR would like to thank M. Gillioz for discussions. This work is supported by MIUR grants PRIN 2017FMJFMW and 2017L5W2PT and INFN grant STRONG.}

\appendix

\section{CFT Collision term}
\label{appA}

In this appendix we derive the collision terms for the Boltzmann equation that governs the distribution of CFT shells. The energy distribution of the CFT can be obtained exploiting the equivalence of CFT shells with a continuum distribution of single particles. We perform this analysis for the case of scalar operators. To achieve we rewrite the space-time density rate $\gamma_0$ as
\begin{eqnarray}
&&\gamma_0\equiv \int_{p_0>|\vec{p}|} \frac{d^4p}{(2\pi)^4 }e^{-\frac {p_0}T}X= \int_0^{\infty} ds\int  \frac{d^3p}{(2\pi)^3 2p_0}  \frac X {2\pi}\nonumber \\
 &&X=   \frac{a_{\op_i}a_{\op_f} }{\pi^2} \frac {d_i-1}{4^{d_i-1} \Gamma(d_i)^2} \frac {d_f-1}{4^{d_f-1} \Gamma(d_f)^2} \left(\frac s {\Lambda^2}\right)^{\Delta} e^{-\frac{p_0}T}
\end{eqnarray}
We recall that for single particles the rate is written as
\begin{equation}
\gamma_0= \frac  1 2 \int \frac{d^3 p}{(2\pi)^3} \frac C E\,,
\end{equation}
where the $C$ is the collision term. This allows us to determine the phase space distribution of single particles produced from the thermal bath (see for example \cite{Bae:2017dpt,Dvorkin:2019zdi}) as
\begin{equation}
f(T,p)=\int_{T}^{T_R} \frac {dT'}{T' H(T')} \frac {C(T',\frac{p\,T'} T)}{\sqrt{M^2+p^2 T'^2/T^2 }}\,.
\end{equation}

The CFT shells are equivalent to a continuum distribution of particles so we define the differential distribution $g(T\,,p\,,s)$
\begin{equation}
  g(T\,,p\,,s)=\int_{T}^{T_R} \frac {dT'}{T' H(T')} \frac {X(T',\frac{p\,T'} T)}{2\pi \sqrt{s+p^2 T'^2/T^2} }\,.
\end{equation}

From this expression, neglecting the dynamics in the CFT sector, we obtain the numerical and energy distribution of the CFT shells by integrating over $p$ and $s$,
\begin{eqnarray}
&&n_{\rm shell}=\frac 1{2\pi^2} \int_0^\infty ds \int_0^\infty dp p^2  g(T,p,s)=a_1 \left(1- \left(\frac T {T_R}\right)^{2\Delta -1}  \right) \frac {M_p}{T_R} \left(\frac {T_R} {\Lambda}\right)^{2\Delta} T^3\,, \nonumber \\
&&a_1 =\frac {3 \sqrt{5/2}}{8\sqrt{g} \pi^6} \frac {\Gamma[d_i+d_f-3]\Gamma[d_i+d_f-2]}{(2d_i+2d_f-9)\Gamma[d_i]\Gamma[d_i-1]\Gamma[d_f]\Gamma[d_f-1]}\,,
 \nonumber \\
&&\rho_{\rm shell}=\frac 1{2\pi^2} \int_0^\infty ds \int_0^\infty dp p^2 \sqrt{s+p^2}  g(T,p,s)=a_2 \left(1- \left(\frac T {T_R}\right)^{2\Delta}  \right)  \left(\frac {T_R} {\Lambda}\right)^{2\Delta} \,M_p T^3 \,,\nonumber \\
&&a_2 =
\frac {3 \sqrt{5/2} }{8\sqrt{g} \pi^6} \frac {\Gamma[d_i+d_f-5/2]\Gamma[d_i+d_f-3/2]}{(d_i+d_f-4)\Gamma[d_i]\Gamma[d_i-1]\Gamma[d_f]\Gamma[d_f-1]}\,,
\end{eqnarray}
where we used
\begin{equation}
I=\int_0^\infty dx\int_0^\infty dy \frac {\exp[\sqrt{-x^2+y^2}]x^{\alpha}y^2}{\sqrt{x^2+y^2}} =2^\alpha\Gamma\left[\frac{\alpha+1}2\right]\Gamma\left[\frac{\alpha+3}2\right]\,. 
\end{equation}

Let us note that the energy density redshifts as matter\footnote{This is consistent with conformal invariance as the shell states are not thermal states.}. 
This follows from the fact that the production is dominated by shells with $\sqrt{s}\sim T$ that become 
immediately non-relativistic. This however does not take into account that the evolution of heavy CFT shells after production.
In weakly coupled scenario  the multiparticle states are not coherent and the constituents evolve as free massless particles.
In the strongly coupled scenario the states are coherent but they are typically unstable towards decay to lower mass shells.

In the weakly coupled regime, the CFT collision term is simply related to the particle collision term. Let us consider the case where the CFT shells are made of 2 massless particles. 
We can think that each shell produced by the operator $\op_f$ decays instantaneously into the final particles. In the centre of mass frame the decay is isotropic. For a shell $(s, \vec{p})$ the distribution of final particles can be simply obtained by boosting from the CM frame distribution
\begin{equation}
\vec{v} =\frac {\vec{p}} {p_0}\\,~~~~~~~~\gamma=\sqrt{\frac{p_0^2}{p_0^2-\vec{p}^2}}
\end{equation}
The energy of a massless particle travelling at an angle $\theta_{\rm CM}$ relative to $\vec{p}$ in the centre of mass frame, in the boosted frame is then
\begin{equation}
E= \frac {1}2(p_0-p x)\,,~~~~~~~-1\le x \le1
\end{equation}
where $x\equiv\cos \theta_{\rm CM}$.
To obtain the collision term of the final state massless particles we change variables $p_0\to E$ so that $dp_0=2 dE$.
Averaging over $\theta_{CM}$ and taking into account that each shell produces 2 massless particles one finds,
\begin{equation}
\begin{split}
\frac C E &=\frac 2 {E^2} \int_{-1}^1 dx \int_0^{\frac{2E}{1-x}} p^2 dp e^{-p_0/T} \frac X {2\pi}\,  \\
&=\frac{a_{\op_i}a_{\op_f}}{8\pi^3} \frac{\Gamma[d_i+d_f-3]}{\Gamma[d_i]\Gamma[d_i-1]\Gamma[d_f]\Gamma[d_f-1]} E^{d_i+d_f-5} e^{-\frac E T}\frac{T^{d_i+d_f-2}}{\Lambda^{2d_i+2d_f-8}}\,.
\end{split}
\end{equation}
For example, in the case of the the Higgs portal this reproduces the result in \cite{Garani:2021zrr}.

The lack of coherence of weakly coupled multiparticle states also affects the thermal distribution. 
For classical statistics the distribution of multiparticles states follows a Maxwell-Boltzmann distribution because $f(E_1)f(E_2)=f(E_1+E_2)$. 
This however is not verified with quantum statistics. For free particles we can find the equilibrium distribution of two particle state by considering
convoluting the thermal distributions of 1 particles states. For 2-particles one finds explicitly,
\begin{equation}
f_{\rm shell}(p_0) =\frac{\int_{-p_{\rm max}}^{p_{\rm max}} dp_- f(E_1) f(E_2)}{\int_{-E_{\rm max}}^{E_{\rm max}} dp_-}\,,~~~~~p_0=E_1+E_2\,,~~~~~~p_-=E_1-E_2\,,~~~~~~~p_{\rm max}= \sqrt{p_0^2-s}
\end{equation}
For boson and fermion pairs one finds,
\begin{eqnarray}
f_{\rm bosons}(p_0)&=&\frac {4 T}{(e^{p_0/T}-1)\sqrt{p_0^2-s}}\log \left[\frac {\sinh \frac {p_0+\sqrt{p_0^2-s}}{4T}}  {\sinh \frac {p_0-\sqrt{p_0^2-s}}{4T}}\right]\,, \nonumber \\
f_{\rm fermions}(p_0)&=&\frac {4 T}{(e^{p_0/T}-1)\sqrt{p_0^2-s}}\log \left[\frac {\cosh \frac {p_0+\sqrt{p_0^2-s}}{4T}}  {\cosh \frac {p_0-\sqrt{p_0^2-s}}{4T}}\right]\,.
\end{eqnarray}
To take into account of quantum statistics thus the thermal averaging of CFT amplitudes should be performed using these distributions.

\pagestyle{plain}
\bibliographystyle{jhep}
\small
\bibliography{biblio}

\end{document}